\documentclass[12pt]{article}
\usepackage{amsfonts,amsmath,amssymb,mathrsfs, comment}

\title{Predictions and Primitive Ontology\\ in Quantum Foundations:\\A Study of Examples}
   
\author{
Valia Allori\footnote{Department of Philosophy, Northern Illinois University,
     Zulauf Hall 920, DeKalb, IL 60115, USA. E-mail: vallori@niu.edu},
Sheldon Goldstein\footnote{Departments of Mathematics, Physics and
     Philosophy, Rutgers University, Hill Center,  
     110 Frelinghuysen Road, Piscataway, NJ 08854-8019, USA.
     E-mail: oldstein@math.rutgers.edu},\\
Roderich Tumulka\footnote{Department of Mathematics,
     Rutgers University, Hill Center,  
     110 Frelinghuysen Road, Piscataway, NJ 08854-8019, USA.
     E-mail: tumulka@math.rutgers.edu},
 and Nino Zangh\`\i\footnote{Dipartimento di Fisica dell'Universit\`a
     di Genova and INFN sezione di Genova, Via Dodecaneso 33, 16146
     Genova, Italy. E-mail: zanghi@ge.infn.it}
}

\date{August 21, 2012}

\newcommand{\Hilbert}{\mathscr{H}}
\newcommand{\conf}{\mathcal{Q}}

\renewcommand{\Im}{\mathrm{Im}}
\newcommand{\EEE}{\mathbb{E}}

\newcommand{\PPP}{\mathbb{P}}
\newcommand{\RRR}{\mathbb{R}}

\newcommand{\scp}[2]{\langle #1|#2 \rangle}
\newcommand{\pr}[1]{| #1 \rangle \langle #1 |}
 
\newcommand{\tr}{\mathrm{tr}}
\newcommand{\sphere}{\mathbb{S}}
\newcommand{\E}{\mathscr{E}} 
\newcommand{\F}{G} 
\newcommand{\GRW}{\mathrm{GRW}}
\newcommand{\Qu}{\mathrm{Qu}}

\newcommand{\sys}{{\mathrm{sys}}}

\newcommand{\env}{{\mathrm{env}}}

\newcommand{\cpm}{\mathscr{C}} 
\newcommand{\dm}{\rho} 

\newcommand{\ket}[1]{| #1 \rangle}
\newcommand{\dead}{\ket{\text{dead}}}
\newcommand{\alive}{\ket{\text{alive}}}
\newcommand{\be}{\begin{equation}}
\newcommand{\ee}{\end{equation}}

\newcommand{\x}[1]{{#1}}

\begin{document}
\maketitle
\begin{abstract}
A major disagreement between different views about the foundations of quantum mechanics concerns whether for a theory to be intelligible as a fundamental physical theory it must involve a ``primitive ontology'' (PO), i.e., variables describing the distribution of matter in 4-dimensional space-time. In this paper, we illustrate the value of having a PO. We do so by focussing on the role that the PO plays for extracting predictions from a given theory and discuss valid and invalid derivations of predictions. To this end, we investigate a number of examples based on toy models built from the elements of familiar interpretations of quantum theory.

\medskip

\end{abstract}

\begin{center}
\textit{Dedicated to Tim Maudlin on the occasion of his 50th birthday}
\end{center}

\newpage
\tableofcontents

\section{Introduction}

This paper is based on the view that any viable interpretation of quantum mechanics, or more generally any fundamental physical theory, must involve variables describing the distribution of matter in space and time. Such variables describe the \emph{primitive ontology} (PO) of the theory \cite{AGTZ06}. Examples of theories with a PO, examples very relevant to this paper, include Bohmian mechanics \cite{Bohm52, Bell66,survey} (with the PO given by the particles), Schr\"odinger's first quantum theory \cite{Schr26,mw} (with the PO given by the mass or charge density), as well as GRWm and GRWf, two versions of the Ghirardi--Rimini--Weber (GRW) theory \cite{GRW86,Bell87} of spontaneous wave function collapse corresponding to two choices of PO: the \emph{matter density ontology} (GRWm) and the \emph{flash ontology} (GRWf). We recall the definitions of GRWm and GRWf in Section~\ref{sec:GRWmf} and that of Bohmian mechanics in Section~\ref{sec:BM1}.

This paper concerns derivations of empirical predictions, i.e., of predictions that can be tested empirically. We believe that, \emph{for a theory with a PO, a derivation of empirical predictions should be based on the PO.} To elucidate this statement is the goal of this paper. We describe how a valid derivation of empirical predictions should work and where it should use the PO and the laws governing it. We do so mainly by means of examples. Some of the examples (Section \ref{sec:rolePO}) are valid derivations found in the literature concerning GRW or other serious theories, but most of them (Section \ref{sec:examples}) are novel and concern toy theories that we have concocted for the purposes of this paper. Some of these toy theories make completely wrong predictions; but that does not preclude them from exemplifying valid derivations of predictions (which just happen to disagree with empirical findings). For a more systematic analysis of why a fundamental physical theory needs to explicitly involve a PO (describing the distribution of matter in space and time), see \cite{AGTZ06,Mau10}.

In the following we simply say ``predictions'' for ``empirical predictions,'' and ``empirical content'' for the sum of all empirical predictions of a theory. We do not, in this paper, compute any specific predictions for specific experiments. For simplicity, we limit our considerations to the non-relativistic quantum mechanics of $N$ spinless particles; the models we describe can easily be modified so as to incorporate spin.

\section{The GRWm and GRWf Theories}
\label{sec:GRWmf}

\x{For introductory presentations of the idea behind theories of spontaneous wave function collapse, such as GRW theory, see \cite{Bell87,Gol98,putnam,Fri07,AGTZ06}.} Detailed introductions to the GRWm and GRWf theories have been given recently in \cite{AGTZ06,Tum06d,grw3A}. Here we give only a brief description. 

\subsection{The GRW Process}
\label{sec:grwprocess}

In both GRWm and GRWf the evolution of the wave function follows, instead of the Schr\"odinger equation, a stochastic jump process in Hilbert space, called the GRW process. Consider a quantum system of (what would normally be called) $N$ ``particles,'' described by a wave function $\psi = \psi(q_1, \ldots,q_N)$, $q_i\in \RRR^3$, $i=1,\dots, N$.  The GRW process behaves as if an ``observer'' outside the universe made unsharp ``quantum measurements'' of the position observable of a randomly selected particle at random times $T_1,T_2,\ldots$ that occur with constant rate $N\lambda$, where $\lambda$ is a new constant of nature of order of $10^{-16}\, \text{s}^{-1}$, called the collapse rate per particle. The wave function ``collapses'' at every time $T=T_k$, i.e., it changes discontinuously and randomly as follows. The post-collapse wave function $\psi_{T+}=\lim_{t\searrow T}\psi_t$ is obtained from the pre-collapse wave function $\psi_{T-}=\lim_{t\nearrow T} \psi_t$ by multiplication by a Gaussian function,
\be\label{collapse}
\psi_{T+}(q_1,\ldots,q_N) = \frac{1}{Z} g(q_I-X)^{1/2}\, \psi_{T-}(q_1,\ldots,q_N) \,,
\ee
\x{where
\be\label{Gaussian}
g(x) = \frac{1}{(2\pi\sigma^2)^{3/2}} e^{-\frac{x^2}{2\sigma^2}}
\ee
is the 3-dimensional Gaussian function of width $\sigma$, $I$ is chosen randomly from $1,\ldots,N$, and
\be\label{Zdef}
Z=Z(X)=\biggl(\int_{\RRR^{3N}} dq_1\cdots dq_N\, g(q_I-X)\, |\psi_{T-}(q_1,\ldots,q_N)|^2\biggr)^{1/2}
\ee
is a normalization factor. The width $\sigma$ is another new constant of nature of order of $10^{-7}\, \text{m}$, while the center $X=X_k$ is chosen randomly with probability density $\rho(x)=Z(x)^2$. We will refer to $(X_k,T_k)$ as the space-time location of the collapse.
}

Between the collapses, the wave function evolves according to the Schr\"odinger equation corresponding to the standard Hamiltonian $H$ governing the system, e.g., given, for $N$ spinless particles, by
\begin{equation}\label{eq:H}
  H=-\sum_{k=1}^N\frac{\hbar^2}{2m_k}\nabla^2_{q_k}+V,
\end{equation}
where $m_k$, $k=1, \ldots, N$, are the masses of the particles, and $V$ is the potential energy function of the system. Due to the stochastic evolution, the wave function $\psi_t$ at time $t$ is random.

\x{This completes our description of the GRW law for the evolution of the wave function. According to GRW theory, the wave function $\psi$ of the universe evolves according to this stochastic law, starting from the initial time (say, the big bang). As a consequence \cite{BG03,grw3A}, a subsystem of the universe (comprising $M<N$ ``particles'') will have a wave function $\varphi$ of its own that evolves according to the appropriate $M$-particle version of the GRW process during the time interval $[t_1,t_2]$, provided that $\psi(t_1)=\varphi(t_1)\otimes\chi(t_1)$ and that the system is isolated from its environment during that interval.

Another remark concerns density matrices. It is a standard fact that with every probability distribution $\mu(d\psi)$ on the unit sphere 
\be
\sphere(\Hilbert)=\bigl\{\psi\in\Hilbert:\|\psi\|=1\bigr\}
\ee
of a Hilbert space $\Hilbert$ there is associated a density matrix
\be
\dm_\mu = \EEE_\mu \pr{\psi} = \int\limits_{\sphere(\Hilbert)} \mu(d\psi) \, \pr{\psi}\,,
\ee
where $\EEE$ means expectation. Since the GRW process is stochastic, $\psi_t$ is random, and with its distribution $\mu_t$ there is associated a density matrix $\dm_t=\dm_{\mu_t}$. It turns out that $\dm_t$ evolves according to an autonomous equation (i.e., one that depends on $\mu_t$ only through $\dm_t$), the ``master equation.'' This equation is of a type known as a Lindblad equation, or quantum dynamical semigroup, and tends to evolve pure states into mixed states. Although the mathematical details play no role in this paper, we give the master equation for the sake of completeness:
\begin{equation}\label{M}
  \frac{d \dm_t}{d t} = -\tfrac{i}{\hbar} [H,\dm_t] 
  + \lambda \sum_{k=1}^N \int d^3 x \, g_{k,x}^{1/2} \, \dm_t \, g_{k,x}^{1/2} - N\lambda \dm_t\,.
\end{equation}
Here, $g_{k,x}^{1/2}$ is the multiplication operator by the function $g(q_k-x)^{1/2}$ with $g$ the 3-dimensional Gaussian function as in \eqref{Gaussian}.
}

\bigskip

We now turn to the primitive ontology (PO). In the subsections below we present two versions of the GRW theory, based on two different choices of the PO, namely the \emph{matter density ontology} (GRWm in Section~\ref{sec:GRWm}) and the \emph{flash ontology} (GRWf in Section~\ref{sec:GRWf}).

\subsection{GRWm}
\label{sec:GRWm}

GRWm postulates that, at every time $t$, matter is continuously distributed in space with density function $m(x,t)$ for every location $x\in \RRR^3$, given by
\begin{align}
 m(x,t) 
 &= \sum_{i=1}^N m_i \int\limits_{\RRR^{3N}}  dq_1 \cdots dq_N \, \delta^3(q_i-x) \,  
 \bigl|\psi_t(q_1, \ldots, q_N)\bigr|^2 \label{mdef}\\
&= \sum_{i=1}^N m_i \int\limits_{\RRR^{3(N-1)}} \!\!\!\! dq_1\cdots dq_{i-1} \, dq_{i+1}\cdots dq_N\,
\bigl| \psi_t(q_1,\ldots,q_{i-1},x,q_{i+1},\ldots,q_N) \bigr|^2\,.
\end{align}
In words, one starts with the $|\psi|^2$--distribution in configuration space $\RRR^{3N}$, then obtains the marginal distribution of the $i$-th degree of freedom $q_i\in \RRR^3$
by integrating out all other variables $q_j$, $j \neq i$, multiplies by the mass associated with $q_i$, and sums over $i$. Alternatively, \eqref{mdef} can be rewritten as
\begin{equation}\label{mdef2}
  m(x,t) = \scp{\psi_t}{\widehat{M}(x) |\psi_t}
\end{equation}
with $\widehat{M}(x) = \sum_{i} m_i \, \delta^3(\widehat{Q}_i - x)$ the mass density operator, \x{defined in terms of the position operators $\widehat{Q}_i \psi(q_1,\ldots,q_N) = q_i\,\psi(q_1,\ldots,q_N)$.}

\subsection{GRWf}
\label{sec:GRWf}

According to  GRWf, the PO is given by ``events'' in space-time called flashes, mathematically described by points in space-time. What this means is that in GRWf matter is neither made of particles following world lines, nor of a continuous distribution of matter such as in GRWm, but rather of discrete points in space-time, in fact finitely many points in every bounded space-time region.

In the GRWf theory, the space-time locations of the flashes can be read off from the history of the wave function: every flash corresponds to one of the spontaneous collapses of the wave function, and its space-time location is just the space-time location of that collapse. The flashes form the set
\be
  F=\{(X_{1},T_{1}), \ldots, (X_{k},T_{k}), \ldots\}
\ee
(with $T_1<T_2<\ldots$). Alternatively, we may postulate that flashes can be of $N$ different types (``colors''), corresponding to the mathematical description
\be
F=\{(X_1,T_1,I_1),\ldots, (X_k,T_k,I_k),\ldots\}\,,
\ee
with $I_k$ the number of the particle affected by the $k$-th collapse.

Note that if the number $N$ of degrees of freedom in the wave function is large, as in the case of a macroscopic object, the number of flashes is also large (if $\lambda=10^{-15}$ s$^{-1}$ and $N=10^{23}$, we obtain $10^{8}$ flashes per second). Therefore, for a reasonable choice of the parameters of the GRWf theory, a cubic centimeter of
solid matter contains more than $10^8$ flashes per second. That is to say that large numbers of flashes can form macroscopic shapes, such as tables and chairs. That is how we find an image of our world in GRWf.

\bigskip

We should remark that the word ``particle'' can be misleading. According to GRWf, there are no particles in the world, just flashes and a wave function. According to GRWm, there are no particles, just continuously distributed matter and a wave function. The word ``particle'' should thus not be taken literally (just like, e.g., the word ``sunrise''); we use it only because it is common terminology in quantum mechanics.

\section{Predictions and Primitive Ontology}
\label{sec:rolePO}

\x{In the PO view, a satisfactory theory should have a PO. In this view, the PO also permits the derivation of predictions. For example,} if we want to derive that in a certain experiment the pointer of the apparatus will end up pointing to the value $z$ with a certain probability then, according to the PO view, we need to derive that the configuration of the PO will be such that the matter of the pointer is in a configuration corresponding to the pointer pointing to $z$. In contrast, it would not be appropriate in this view to merely show that the wave function lies (approximately) in a subspace of Hilbert space corresponding to the pointer pointing to $z$. 

\x{For example, a number of empirical predictions of GRW theory have been derived in \cite{GRW86,PS94, Leg02, JPR04, Adl07}. It was found that the predictions deviate from those of quantum mechanics but only so slightly that no experimental test has been possible so far \cite{Adl07,FT11}. However, the logical clarity of the derivations in \cite{GRW86,PS94, Leg02, JPR04, Adl07} leaves something to be desired, as they do not refer to the PO but limit themselves to analyses of the wave function. While we do not dispute that the claimed predictions are indeed predictions of GRWm and GRWf, we do see a gap in the derivation. The situation is similar to that of a calculation that yields the correct result but is not mathematically rigorous. Here, the problem is not one of mathematical rigor but of clarity---philosophical, ontological, conceptual, and physical clarity. We will describe in this chapter, particularly in Section~\ref{sec:mainthm}, how to close this gap.
}

\subsection{Calibration Functions}

\x{Given an experiment $\E$, its} outcome $Z$ is a function of the (configuration of the) PO,
\begin{equation}\label{ZPO}
  Z = \zeta(PO)\,.
\end{equation}
That is, in GRWf $Z=\zeta_\mathrm{GRWf}(F)$, and in GRWm $Z=\zeta_\mathrm{GRWm}(m)$. Similarly, the PO plays a key role for the claim of empirical equivalence between two theories (for which there will be several examples in this paper). 

Let us elaborate a bit on what the function $\zeta$ should look like. In GRWm, it is natural that $Z$ should be a functional of $m(\cdot,t)$, the distribution of matter at the time $t$ when the experiment is completed. What the $\zeta$ function does is essentially to read off the outcome from the display of the apparatus. For example, if the outcome is displayed by means of the position of a needle on a scale, $\zeta$ should read off from $m$ the position of the needle, and may concretely be the following: Suppose the experiment is so arranged that the region $R\subseteq \RRR^3$, for simplicity a cuboid $R=[a_1,b_1] \times [a_2,b_2] \times [a_3,b_3]$, contains no other matter than the needle, and suppose the scale is along the $x_1$-axis between $a_1$ and $b_1$. Then the mean $x_1$ coordinate of the matter distribution at time $t$ inside $R$ is given by
\be
\langle x_1 \rangle = \frac{\int_R dx_1\, dx_2 \, dx_3\, x_1 \, m(x_1,x_2,x_3,t)}
{\int_R dx_1\, dx_2 \, dx_3 \, m(x_1,x_2,x_3,t)}\,,
\ee
and a typical choice of calibration function would be
\be\label{zetameanx1}
\zeta(m) = z_0 + \alpha \langle x_1 \rangle
\ee
with suitable proportionality constant $\alpha$. Note that $\zeta$ is a functional of $m(\cdot,t)$. If we require that $Z$ be a discrete variable, we may replace \eqref{zetameanx1} with a suitable step function, such as 
\be\label{zetameanx1discrete}
\zeta(m) = z_0 + [\alpha \langle x_1 \rangle]\,,
\ee
where $[z]$ denotes the nearest integer to the real number $z$.

In GRWf, $Z$ must depend on the history in an entire time interval, say $[t,t+\Delta t]$ with $t$ the time when the experiment is completed and $\Delta t$, say, a millisecond. In the example of the needle pointing to the outcome, $\zeta$ needs to read off the position of the needle from the flashes and rescale it appropriately. As a concrete example, suppose again the region $R$ contains no other matter than the needle, and suppose the scale is along the $x_1$-axis between $a_1$ and $b_1$. Then the mean $x_1$ coordinate of the flashes during $[t,t+\Delta t]$ in $R$ is given by
\be
\langle x_1 \rangle = \frac{\sum_k 1_{t\leq T_k\leq t+\Delta t} \, 1_{X_k \in R} \, (X_k)_1}
{\sum_k 1_{t\leq T_k\leq t+\Delta t} \, 1_{X_k \in R}}\,,
\ee 
where $1_C$ is $1$ when the condition $C$ is satisfied and 0 otherwise. A typical choice of calibration function would be given in terms of this $\langle x_1 \rangle$ by \eqref{zetameanx1} or \eqref{zetameanx1discrete}.

\subsection{Taking the PO Seriously}
\label{sec:seriousPO}

As a consequence of the view that the PO represents matter, we are forced to take the PO seriously. 

For example, the transformation behavior of the PO under a Lorentz (or Galilean) transformation is
constrained by its geometrical nature, and that of the non-primitive ontology (i.e., the remaining part of the ontology) is constrained by its relation to the PO; see section 4.2 of \cite{AGTZ06} for elaboration. 
If the PO consists of flashes, then the flashes have to transform like space-time points under Lorentz transformations. If the PO consists of particle world lines, then they have to transform like world lines, i.e., by transforming every space-time point on the world line. Put differently, a world line must be an unambiguous set of space-time points, independently of the choice of coordinates. 

As another example for what it means to take the PO seriously, we look at a difficulty that arose in \cite{Bas07} \x{with} an over-simplified discussion of how to derive in GRWm that, in a suitable situation, a pointer is pointing to $z$. \x{The} coordinates $x_1, \ldots, x_N \in \RRR^3$ of the configuration space $\RRR^{3N}$ of a pointer consisting of $N$ ``particles''  \x{were decomposed} into the center of mass $x_{\text{cm}} = \sum_i m_i x_i/\sum_j m_j \in \RRR^3$ and relative coordinates $r_1,\ldots, r_{3N-3}$, and \x{it was} derived that in this decomposition the wave function, in a suitable situation, is approximately a product $\psi_{\text{cm}}(x_\text{cm}) \, \psi_{\text{rel}}(r_1, \ldots, r_{3N-3})$, where the first factor, the center-of-mass wave function $\psi_\text{cm}$, is a very narrow wave packet. \x{Therefore, it was suggested,} the matter density associated with the center-of-mass, $m_\text{cm}(x,t) = |\psi_\text{cm}(x,t)|^2$, is very narrow, too (thus making the position of the pointer sufficiently precisely defined). The difficulty with this argument is that $m_\text{cm}$ is not the right quantity to look at: It is not the PO, not the matter, not real; it is just a mathematical quantity. What counts is whether $m$ as defined in \eqref{mdef} is concentrated in the right location, and there is no simple relation between $m$ and $m_\text{cm}$. For example, the width of $m_\text{cm}$ is, in realistic examples, $10^{-13}$~m whereas that of $m$ is $10^{-3}$~m (= the width of the pointer). So \x{this} argument did not take the PO seriously. \x{Rather, it still treated $\psi$ in a more or less conventional way as providing probabilities for configurations of particles, as that is the situation in which the extreme narrowness of $\psi_\text{cm}$ can be relevant.}

The issue of taking the PO seriously also arises in the context of studying the limitations on knowledge in GRWm and GRWf, a topic we plan to discuss in detail in a future work \cite{knowledge2} (and that we have outlined in \cite{grw3A}): Inhabitants of a GRWf or GRWm world cannot measure the times and locations of the collapses, although these values are well defined according to these theories. That is, some things that are real cannot always be measured with arbitrary accuracy. But if something cannot be measured, one may be tempted to not take it seriously. So, if there are limitations to measuring the variables representing the PO, one may be tempted not to take the PO seriously. Needless to say, we recommend resisting this temptation. The conclusion ``unobservable, therefore unreal'' is, of course, not a good one.

\bigskip

Another remark concerns the fine difference between the names ``matter density'' and ``mass density'' for the $m$ function. The fact that its definition \eqref{mdef} involves the quantity $m_i$ usually called ``the mass of particle $i$'' suggests the name ``mass density'' for $m$, which, as we shall argue, can be misleading. We prefer the name ``matter density'' because it reflects what we take to be the fundamental meaning of the $m$ function. Let us explain.

Suppose we had postulated, instead of \eqref{mdef}, the following formula for the $m$ function:
\begin{equation}\label{mdefe}
 m(x,t) = \sum_{i=1}^N e_i \int\limits_{\RRR^{3N}}  dq_1 \cdots dq_N \, \delta(q_i-x) \,  \bigl|\psi_t(q_1, \ldots, q_N)\bigr|^2 \,.
\end{equation}
This is the same equation as \eqref{mdef}, but with $m_i$, the ``mass of particle $i$,'' replaced by $e_i$, the ``charge of particle $i$.'' In this case, it would evidently not be appropriate any more to call the $m$ function the mass density; rather ``charge density'' would seem more appropriate. At the same time, we may postulate that the $m$ function still represents the \emph{matter density}---i.e., we may postulate that matter is distributed continuously with density $m$---and regard \eqref{mdefe} as merely a different law for how matter is distributed.\footnote{It may be a difficulty with this postulate that \eqref{mdefe} can be negative, while it is normal to think of matter density as non-negative. This difficulty might be addressed by adding a sufficiently large positive constant on the right hand side, or perhaps just by insisting that matter density can be negative, for example representing some kind of ``anti-matter'' (not necessarily related to the usual meaning of this word). For the purpose of this section, readers may ignore this difficulty.} In that case, the meaning of the $m$ function would not have changed in any way---we would only have changed the law governing it. That is why this meaning is better conveyed by the name ``matter density.'' Moreover, the matter that we postulate in GRWm and whose density is given by the $m$ function does not ipso facto have any such properties as mass or charge; it can only assume various levels of density. For example, the $m$ function is not a source of an electromagnetic field.

In addition, the terms ``mass density'' and ``charge density'' may easily suggest that both quantities physically coexist: Is it not natural to expect that an extended object possesses both a mass density and a charge density (different from each other)? In fact, it seems completely coherent (though perhaps not desirable) to postulate that the PO involves both a mass density and a charge density (different from each other). If, in contrast, we postulate that the PO consists of only one density---the matter density---and choose either \eqref{mdef} or \eqref{mdefe} to determine it, then only one of the two functions represents something real (more precisely, represents the PO), whereas the other is a pure mathematical fiction. In other words, in GRWm with \eqref{mdef} as the law of $m$, the formula on the right hand side of \eqref{mdefe} lacks any physical significance, just as it does in GRWf.

\subsection{Examples From the Literature}

We now look at earlier arguments based on the connection we are discussing between empirical predictions and PO. These are also examples of valid derivations of empirical predictions that do not not suffer from the gap we complained about (in the second paragraph of this chapter).

\begin{enumerate}
\item The first example is a proof of no-signaling (i.e., the impossibility of transmitting messages faster than light between two distant observers, each acting on and observing one of two entangled quantum objects) in GRWf due to Bell \cite{Bell87}. The proof shows that for two non-interacting systems (here, the system can be taken to be one object together with a nearby apparatus), the marginal distribution of the flashes pertaining to system 1 depends on the entangled wave function only through its reduced density matrix (with system 2 traced out); nor does it depend on the Hamiltonian of system 2. Thus, it does not depend upon any message that observer 2 may wish to transmit. Now, \emph{since outcomes of experiments are functions of the flashes}, the outcome that observer 1 sees cannot depend on the message observer 2 may have wished to transmit, qed. In Bell's words \cite{Bell87},
\begin{quotation}
Events in one system, considered separately, allow no inference about [\ldots] external fields at work in the other, \ldots\ nor even about the very existence of the other system. There are no ``messages'' in one system from the other. The inexplicable correlations of quantum mechanics do not give rise to signalling between noninteracting systems.
\end{quotation}
A similar proof was given in \cite{Tum04,Tum06c} for a relativistic version of GRWf.

\item The second example is provided by derivations of empirical predictions from Bohmian mechanics, which can be found in many papers, e.g., \cite{Bohm52,Bell80,DGZ04,AZ05}. The PO of Bohmian mechanics consists of particles and their trajectories, and the outcomes of experiments are read off from the particle trajectories, and not directly from the wave function. The empirical content of Bohmian mechanics agrees exactly with that of the quantum formalism ``whenever the latter is unambiguous.'' 

\item The third example is taken from the study of superselection rules by Colin \emph{et al.}~\cite{CDT05}. For superselection rules it is crucial that some self-adjoint operators are \emph{not} observables. In the terminology of Colin \emph{et al.}, a ``weak superselection rule'' means that no experiment can distinguish between a \emph{superposition} of vectors from different superselection sectors in Hilbert space and a suitable \emph{mixture} thereof. \x{This occurs only if all observables commute with all projections to superselection sectors, because otherwise a quantum measurement of such an observable would distinguish between them.} Colin \emph{et al.}\ proved certain weak superselection rules for Bohmian mechanics (and Bohmian versions of quantum field theory) as well as for GRWm and GRWf, and the proof uses that the outcome of any experiment is a function of the PO (and not of the wave function). Indeed, while the PO does not permit us to distinguish between a superposition and a mixture of contributions from different sectors, the wave function would trivially permit this.

In more detail, Colin \emph{et al.}\ first showed that (in certain situations) the distribution of the flashes in GRWf does not distinguish between a superposition and a suitable mixture (``strong superselection''), and concluded from this that the distribution of outcomes of experiments does not distinguish between them either. Due to the empirical equivalence between GRWf and GRWm, the same experiment does not distinguish between them in GRWm. To sum up, the proof that some operators are \emph{not} observables used that results of observations must be read off from the PO, and then proceeded with a suitable analysis of the PO.

\item Colin and Struyve discussed in \cite{CS07} whether their Dirac sea model is empirically equivalent with orthodox quantum field theory. They suggested an unusual PO containing a huge (or even infinite) number of particles with trajectories, one for every electron in the Dirac sea. For the theory to make the right predictions, we need that macroscopic facts such as outcomes of experiments can be read off from this PO---and that is what Colin and Struyve analyzed.

\item In \cite{grw3A} we have shown (and outlined before in \cite{AGTZ06}) that GRWm and GRWf are empirically equivalent. This is a non-trivial statement if and only if the PO is taken seriously.

\item Feldmann and Tumulka \cite{FT11} considered other values for $\sigma$ and $\lambda$ than suggested originally \cite{GRW86} ($\sigma=10^{-7}$ m, $\lambda=10^{-16}$ s$^{-1}$) and evaluated for which points in the $\sigma\lambda$-plane the GRWm and GRWf theories are empirically adequate, and for which they are philosophically satisfactory. Their criterion for being philosophically satisfactory is that the PO looks macroscopically like what humans normally imagine reality to be like. As a consequence, for determining the philosophically satisfactory region they need to pay attention to the behavior of the PO.

\item In \cite{grw3A} we have derived a measurement formalism for GRWm and GRWf, i.e., a set of rules for computing the predictions of GRWm and GRWf for any experiment. We call it the \emph{GRW formalism}; it is spelled out in Section~\ref{sec:GRWformalism} below. The GRW formalism is analogous to the \emph{quantum formalism}, i.e., the usual set of rules for making predictions in quantum mechanics. Both the quantum and the GRW formalism can be formulated as three rules, the first specifying how an isolated system evolves, the second specifying the probability distribution of the outcome of a given experiment, and the third specifying the change on the system's state depending of the outcome obtained. That is, the third rule is a ``collapse rule,'' concerning, however, not spontaneous collapses but the collapse induced by the intervention of a macroscopic apparatus. Put differently, the GRW formalism is what the quantum formalism needs to be replaced with if we pay attention to the small deviations from the quantum formalism predicted by GRWm and GRWf. 

The derivation of the GRW formalism in \cite{grw3A} takes the PO seriously. We discuss it to some extent in the next two subsections.
\end{enumerate}

\subsection{The Main Theorem About Operators in The GRW Formalism}
\label{sec:mainthm}

Before talking about the full GRW formalism, we begin with a statement that is part of the GRW formalism, the \emph{main theorem about operators: With every experiment $\E$ on a system and every possible outcome $z$ of $\E$ there is associated an operator $P_z$ acting on the Hilbert space $\Hilbert_\sys$ of the system. When $\E$ is performed on a system with wave function $\psi$, the outcome $Z$ is random with probability distribution}
\be\label{PPPpsi}
\PPP(Z=z) = \scp{\psi}{P_z|\psi}\,.
\ee
This statement is actually true in both the quantum formalism and the GRW formalism; however, to the same experiment, the GRW formalism may assign operators $P_z^\GRW$ different from the $P_z^\Qu$ assigned by the quantum formalism. Formulas for $P_z^\Qu$ and $P_z^\GRW$ are given in \cite{grw3A}. (We assume that every experiment has a discrete set of outcomes $z$.)

The operator $P_z$ in \eqref{PPPpsi} may be a projection, but in general it is merely a positive operator (even in the quantum formalism). If the $P_z$ are projections for all $z$, and if the $z$ are real numbers, then the family $\{P_z\}$ corresponds to the self-adjoint operator $A=\sum_z z\,P_z$, usually (but misleadingly) called ``the observable measured'' by $\E$. The $z$ are the eigenvalues of $A$, and the $P_z$ the projections to the eigenspaces. When the $P_z$ are not projections, the family $\{P_z\}$ forms a \emph{positive-operator-valued measure (POVM)} (meaning, in this discrete case, that the positive operators $P_z$ are such that $\sum_z P_z =I$, the identity operator), a concept well 
known in quantum information theory.

\bigskip

As an example of a valid derivation of predictions that takes the PO seriously, we now outline the derivation of the main theorem about operators from GRWf, following \cite{grw3A}. The key point is that we take the outcome $Z$ of the experiment to be a function of the pattern $F=\{(X_1,T_1,I_1),\ldots, (X_k,T_k,I_k),\ldots\}$ of flashes,
\begin{equation}\label{zetaF}
Z=\zeta(F)\,.
\end{equation}
(It would even be realistic to assume that $Z$ depends only on the flashes of the apparatus during a short time interval appropriate for reading off the outcome, and not those of the system or of the rest of the world, and not those at earlier or later times; but this restriction is not needed here.) 

It is a known fact \cite{Tum07} that the joint distribution of all flashes after time $t$, conditional on all flashes up to $t$, depends quadratically on $\Psi_t$, the wave function of the universe at time $t$. Explicitly, it is given by 
\be\label{flashPOVM}
\PPP(F_{>t}\in S|F_{\leq t}) = \scp{\Psi_t}{\F_t(S)|\Psi_t}
\ee
with $S$ any set of flash histories after $t$ and $\F_t(\cdot)$ a suitable POVM on the space of all flash histories.\footnote{On a continuous space $\Omega$, a POVM associates by definition a positive operator $\F(S)$ with every (measurable) subset $S\subseteq \Omega$ in such a way that $\F(\Omega)=I$ and $\F(S_1\cup S_2)=\F(S_1)+\F(S_2)$ when $S_1\cap S_2=\emptyset$ (and likewise for countable families of pairwise disjoint sets). For discrete $\Omega$, $\F(S)$ can be expressed as $\sum_{z\in S} P_z$ with $P_z=\F(\{z\})$.} 

Let $t$ be the time at which the experiment begins.
Consider splitting the universe into a system (the object of the experiment) and its environment (the rest of the world, including all relevant apparatuses of the experiment), corresponding to a splitting of the Hilbert space into $\Hilbert=\Hilbert_\sys \otimes \Hilbert_\env$. 
We assume independence between the  
system and the environment immediately before $t$, so that
\begin{equation}\label{factorize}
   \Psi_t = \psi \otimes \phi\,.
\end{equation}
Here we regard $\phi$ as fixed, while $\psi$, the initial state of the system upon which  
the experiment is performed, is allowed to vary in the system   
Hilbert space $\Hilbert_\sys$. \x{We may think of $\phi$ as part of the characterization of the  
experiment, although in practice a repetition of the experiment will not begin with exactly the same wave function $\phi$ of the apparatus.}

Therefore, the distribution of the random outcome $Z$ is given by
\begin{equation}\label{PPPZ1}
  \PPP(Z =z) = \PPP\bigl(F \in \zeta^{-1}(z)\bigr) = 
  \scp{\Psi_t}{\F_t \bigl( \zeta^{-1}(z)\bigr)| \Psi_t}
  = \scp{\psi}{P^\GRW_z |\psi}\,,
\end{equation}
where the first scalar product is taken in the Hilbert space of the universe and the second in $\Hilbert_{\sys}$, and $P^\GRW_z$ is the POVM given by
\begin{equation}\label{Efdef}
  P^\GRW_z= \scp{\phi}{\F_f \bigl( \zeta^{-1}(z)\bigr)| \phi}\,,
\end{equation}
where the scalar product is a partial scalar product in the Hilbert space of the environment. Thus, for every experiment in GRWf, the distribution of outcomes is given by a POVM $P^\GRW_z$, which is what we wanted to show.

\subsection{The GRW Formalism}
\label{sec:GRWformalism}

The full GRW formalism is best formulated in terms of density matrices.\footnote{That is because we allow the system under study, system 1, to be entangled with another system, system 2, which does not interact with either system 1 or the apparatus of the experiment; thus, $\dm=\tr_2 \pr{\psi}$, where $\tr_2$ denotes the partial trace and $\psi$ the joint wave function of systems 1 and 2. Since $\psi$ is usually not precisely known, it is also often convenient to take it to be random; then $\dm=\EEE\,\tr_2 \pr{\psi}$.} Perhaps the most remarkable fact about the GRW formalism is that its abstract structure is identical to that of the quantum formalism. It consists of three rules, the first saying how the density matrix $\dm_t$ of a system evolves when the system is isolated (or, when ``the observer'' is not ``taking measurements''); the second (a form of the main theorem about operators) saying that with every experiment (or ``measurement'') there are associated operators, and that the probability for obtaining a particular outcome is given by a trace formula involving $\dm_t$; and the third saying how $\dm_t$ should be changed after the experiment, depending on the outcome. Here are the three rules, in both the quantum and the GRW version:

\begin{itemize}
\item[(i)] A system isolated from its environment has at every time $t$ a density matrix $\dm_t$ which evolves in the quantum formalism according to the unitary (Schr\"odinger) evolution, \x{which for a density matrix reads
\begin{equation}\label{SchrDM}
  \frac{d\dm_t}{dt} = -\tfrac{i}{\hbar} [H, \dm_t]
\end{equation}
with $H$ the system's Hamiltonian, and in the GRW formalism according to the master equation \eqref{M}.}

\item[(ii)] With every experiment $\E$ on a system and every possible outcome $z$ of $\E$ there is associated a positive operator $P_z$ acting on the Hilbert space $\Hilbert_\sys$ of the system. When $\E$ is performed on a system that has density matrix $\dm$ at the beginning of $\E$, the outcome $Z$ is random with probability distribution
\begin{equation}\label{PPP}
  \PPP(Z=z) = \tr \bigl(\dm\, P_z \bigr)\,.
\end{equation}

\item[(iii)]
When the outcome $Z=z$, the density matrix $\dm$ of the system gets replaced by
\begin{equation}\label{C}
  \dm' = \frac{\cpm_z(\dm)}
  {\tr\, \cpm_z(\dm)}\,.
\end{equation}
with $\cpm_z$ a (completely positive) linear operation on density matrices. 

In the quantum formalism for an ideal quantum measurement, the $P_z$ are projections, and
\be
\cpm_z (\dm) = P_z \, \dm \, P_z\,.
\ee
Again, the operations $\cpm_z^\Qu$ provided by quantum formalism and $\cpm_z^\GRW$ by the GRW formalism may differ. Formulas for $\cpm_z^\Qu$ and $\cpm_z^\GRW$ are given in \cite{grw3A}.
\end{itemize}

\section{A Set of Examples}
\label{sec:examples}

A new and useful perspective on GRW theories arises from contrasting them with other theories, even unreasonable ones. For this purpose we develop in this section a set of example theories which we obtain by combining elements of the known theories GRWm, GRWf, and Bohmian mechanics in new, sometimes playful, ways. Some of the theories obtained in this way make completely wrong predictions but are instructive nonetheless since they illustrate the way in which predictions follow from a theory. Others make predictions in agreement with known empirical facts, yet nobody would seriously propose them as fundamental physical theories; still, they allow for illuminating comparisons with GRWm and GRWf.

Here is a ``theory construction kit.'' Choose one of the three primitive ontologies of GRWm, GRWf, and Bohmian mechanics: continuous matter density, flashes, or particles with trajectories. \x{Choose whether the laws governing the PO should involve a wave function $\psi$ or a density matrix $\dm$. Then choose an evolution law for $\psi$ or $\dm$, e.g., the unitary Schr\"odinger evolution, the stochastic GRW evolution for $\psi$, or the GRW master equation \eqref{M} for $\dm$. Then consider simple laws for how $\psi$ (or $\dm$, respectively) may govern the PO.} In this way we arrive at about ten new theories.

\subsection{Bohmian mechanics}
\label{sec:BM1}

To review the elements used in the theory construction kit, let us briefly recall the laws of Bohmian mechanics. Bohmian mechanics is a (non-relativistic) theory of particles in motion. The motion of a system of $N$ particles is provided by their world lines $t \mapsto Q_i(t)$, $i=1, \ldots, N$, where $Q_i(t)$ denotes the position in $\RRR^3$ of the $i$-th particle at time $t$. These world lines are determined by Bohm's law of motion \cite{Bohm52, Bell66,survey}, 
\begin{equation}\label{Bohm}
  \frac{dQ_i}{dt}=v_i^{\psi}(Q_1, \ldots, Q_N)=\frac{\hbar}{m_i}
  \Im \frac{\psi^{*}\nabla_i \psi}{\psi^{*}\psi}(Q_1\ldots,Q_N), 
\end{equation}
where the wave function $\psi$ evolves according to Schr\"odinger's equation
\begin{equation}\label{Schr}
 i\hbar\frac{\partial \psi}{\partial t} = H\psi \,,
\end{equation} 
with $H$ the usual nonrelativistic Schr\"odinger Hamiltonian; for spinless particles it is of the form \eqref{eq:H}.

An important probability distribution in Bohmian mechanics is the \emph{quantum equilibrium distribution}
\begin{equation}\label{BMQE}
  p^\psi(q) = |\psi(q)|^2 \,.
\end{equation}
(While the distribution density is usually denoted $\rho$, we write $p$ here in order to reserve the letter $\rho$ for density matrices.)
As a consequence of Bohm's law of motion \eqref{Bohm} and Schr\"odinger's equation \eqref{Schr}, 
$|\psi|^2$ is \emph{equivariant}.  This means that if the configuration $Q(t) = (Q_1(t), \ldots, Q_N(t))$ of a system is random with distribution $|\psi_t|^2$ at some time $t$, then this will be true also for any other time $t$. Because of equivariance, Bohmian mechanics reproduces the predictions of the quantum formalism for typical initial conditions of the universe, as discussed in detail in \cite{DGZ92,DGZ04}.

\subsection{Bohmian Trajectories and GRW Collapses}

We begin with the particle ontology, with the particle trajectories governed by Bohm's law of motion. We consider several ways of combining this with the GRW evolution of the wave function or a similar one.

\subsubsection{Bohm's Law and GRW's Law}

First, suppose simply that the particles move according to the usual Bohmian law of motion \eqref{Bohm}, but that $\psi=\psi_t$ is the GRW wave function, so that the GRW process replaces the unitary evolution. In this theory, which we denote GRWp$_1$, the $|\psi|^2$ distribution for the configuration is not equivariant. 

A world governed by this theory GRWp$_1$ has little resemblance with our world: It behaves in a very unstable way. For example, a system with the wave function of Schr\"odinger's cat has, even before the collapse into either $\dead$ or $\alive$, a configuration of either a dead or a live cat---but the collapse need not agree with that configuration. The cat could be alive before the collapse (i.e., its particle configuration is that of a live cat), and still the collapse could reduce $\psi$ to a state vector close to $\dead$---\x{with however no immediate change in the particle configuration of the cat.}

At this point, the reader may feel unsure whether to conclude that \x{just after collapse} the cat is really alive or that the cat is really dead. That is, partly, what makes this theory philosophically useful, despite its empirical inadequacy: It nicely illustrates the role of the PO. Taking the PO seriously, we must conclude that the cat is really alive; after all, the PO in this theory consists of particles, and the particle configuration is one of a live cat. This illustrates that the mere fact that the wave function is one of a dead cat does not, in and of itself, mean that there is a dead cat.

From that fateful collapse onwards, the configuration is guided, in the sense of Bohm's law of motion \eqref{Bohm}, by that tiny part of $\psi$ that remains of $\alive$ after the collapse, and who knows what happens then. For sure, the further behavior of the configuration will be catastrophic. The further evolution of the configuration is not simply that of a live cat, but will be disturbed by two factors: first, by the fact that the Gaussian collapse factor will change the shape of $\alive$, and second, by the fact that \x{tails of $\dead$, which reach the support of $\alive$ under the Schr\"odinger evolution,} will dominate over the contribution from $\alive$.

\bigskip

\textit{What this example illustrates:} First and foremost, this example illustrates how the same wave function---the GRW wave function---can be combined with a different PO than usual, and thus helps us to get used to the distinction between the wave function and the PO. \x{Second, it illustrates how the particle ontology can be combined with different laws for the wave function---the GRW law instead of the Schr\"odinger equation as in Bohmian mechanics. Third, the example} illustrates what it means to derive predictions from the PO rather than from the wave function, as the wave function is well behaved but the particle configuration is not. In particular, GRWp$_1$ forces us to face the question: Do the predictions follow from the wave function or from the PO? Fourth and finally, GRWp$_1$ shows that the GRW wave function can be part of a theory making completely different predictions than GRWm and GRWf.

\bigskip

Another observation: GRWp$_1$ is presumably an example of a theory \emph{without a formalism}. Note first that the $|\psi|^2$ distribution is not equivariant in GRWp$_1$, in fact there is no equivariant density formula at all. This undercuts the reasons for assuming the initial configuration was $|\psi|^2$ distributed, so that we lose the basis for deriving predictions at all. But even if we postulated the $|\psi|^2$ distribution at some point in time (e.g., the big bang), so that the theory would make unambiguous predictions, the distribution of the outcome $Z$ of an experiment will not be given by a POVM on $\Hilbert_{\sys}$, presumably not even approximately, and we see no reason why the empirical contents could be summarized by a formalism at all.

\subsubsection{Bohm's Law and a Modified GRW Law}

To improve GRWp$_1$, one may think of modifying it a bit: Instead of choosing the collapse center $X$ at random, as prescribed by the GRW process, one could take 
\begin{equation}\label{XQIT}
X=Q_I(T)\,,
\end{equation}
so that the collapse is centered at the actual position of the corresponding particle at the time of the collapse. In other words, for every collapse the time $T$ and the label $I$ are chosen at random as in the GRW process, but the position of the (center of the) collapse is not, but is taken from the particle configuration instead.

Let us call this model GRWp$_2$; it was called GRWp in \cite{AGTZ06}. The behavior of a GRWp$_2$ world is less catastrophic than that of a GRWp$_1$ world, but still the $|\psi|^2$ distribution is not equivariant, and so probabilities cannot be expected to agree with $|\psi|^2$. As with GRWp$_1$, one could consider GRWp$_2$ with $|\psi|^2$ distribution at the big bang, but as with GRWp$_1$, one presumably obtains no POVM, and no formalism.

\bigskip

\textit{What this example illustrates:} \x{The particle ontology can be combined with a multitude of possible laws for the wave function,} each of which is simple and respects the symmetries of the GRW process (invariance under rotations, translations, time translations, and Galilean boosts).

\bigskip

With a little modification in its defining equations, GRWp$_2$ becomes a better behaved theory GRWp$_3$ \cite{Bed,Tum11b}: Instead of \eqref{XQIT}, take the collapse center $X$ to be
\begin{equation}\label{XQITZ}
X = Q_I(T) + Z\,,
\end{equation}
where $Z$ is a random 3-vector that is chosen independently of the past with a Gaussian distribution with mean 0 and covariance matrix diag$(\sigma^2,\sigma^2,\sigma^2)$. It then follows \cite{Tum11b} that the conditional distribution of $Q(t)$, given the $X,I,T$ for all collapses up to time $t$ (or given $\psi_t$), equals $|\psi_t|^2$; that the joint distribution of the $\psi_t$ for all $t\geq 0$ is the same as for the GRW process; and that this theory is empirically equivalent to GRWm and GRWf.

\bigskip

\textit{What this example illustrates:} The empirical content (i.e., the sum of the empirical predictions) of the GRWf and GRWm theories can as well be obtained with the particle ontology, not only with the flash and matter density ontologies.

\subsubsection{Trajectories From the GRW Wave Function}

For the next theory, GRWp$_4$, let us return to the GRW law for $\psi_t$, and consider a particle ontology with positions given by the wave function by means of the law
\be\label{QpsiGRWp4}
Q_i(t)=\scp{\psi_t}{\hat{Q}_i|\psi_t}\,,
\ee
where $\hat{Q}_i$ is the position operator of particle number $i$. That is, the actual position $Q_i(t)$ is what would in orthodox quantum mechanics be the \emph{average} position of particle $i$ (if measured). In contrast to GRWp$_1$ and GRWp$_2$ (and Bohmian mechanics), this theory does not require any initial data about the particle configuration, as the configuration is a function of $\psi_t$. Like GRWp$_1$ and GRWp$_2$, this theory GRWp$_4$ is not empirically adequate. 

This is best seen when considering a system of identical particles, whose wave function obeys permutation symmetry or anti-symmetry. As a preparation, we note that the GRW process as we have defined it in Section~\ref{sec:grwprocess} above does not preserve permutation symmetry. For identical particles the GRW process needs to be defined differently to include a collapse mechanism which preserves permutation symmetry; for the technical details (which will not matter for our purposes) see \cite{Tum05}. So consider the theory defined by that process together with \eqref{QpsiGRWp4}. For a wave function with appropriate permutation symmetry, \eqref{QpsiGRWp4} implies that all particles (of the same species) have the same position. However, it is empirically incorrect that all electrons have the same position.

\bigskip

\textit{What this example illustrates:} Apart from being an example of the multitude of possible laws for the PO, an interesting trait of this theory is that the configuration of the PO supervenes on the wave function by means of the law \eqref{QpsiGRWp4}, as it does in GRWm and GRWf. GRWp$_4$ (or the version of it for identical particles) shows that among the many ways in which a configuration of matter (be it a particle configuration, a continuous matter distribution, or flashes) can supervene on the wave function, different possibilities may strongly disagree about the empirical predictions.

\subsubsection{Configuration Jumps and GRW Law}

Another theory, GRWp$_5$, has the following laws: The wave function $\psi_t$ follows the GRW process, and the configuration moves according to Bohm's law of motion \eqref{Bohm} between the GRW collapses. However, at the time $T$ when $\psi$ collapses around $X\in\RRR^3$ with label $I$, also the configuration $Q$ jumps; more precisely, only the $I$-th particle jumps, and it jumps to the random center $X$ of the GRW collapse:
\begin{equation}\label{GRWp5}
Q_I(T+) = X\,.
\end{equation}

Again, $|\psi|^2$ is not equivariant, and indeed, the behavior of the particles is quite catastrophic: If $\psi(T-)$ is the wave function of Schr\"odinger's cat, the configuration $Q(T-)$ is that of a live cat, and the wave function collapses to that of a dead cat, it may do so with just a few collapses connected to a few particle labels, corresponding to particles that would have to be in different positions depending on whether the cat is dead or alive. As a consequence, the configuration after these few collapses will be one of a live cat, with a few particles moved to where they would have to be if the cat were dead. So, this configuration is very different from what one would normally associate with $\dead$. Also, this configuration may be well outside the support of both $\dead$ and $\alive$; it may be a configuration for which $|\psi|^2$ is literally zero, or much smaller than even the remains of $\alive$ after the collapse. And the behavior of such configurations should be expected to be catastrophic.

\bigskip

\textit{What this example illustrates:} 
One normally thinks that when $\psi=\dead$ then there is a dead cat. However, as GRWp$_5$ illustrates, in the PO view this cannot be taken for granted but must be checked.

\subsubsection{Another Way of Configuration Jumps and GRW Law}

\x{Within the framework that $\psi_t$ follows the GRW process and the configuration moves according to Bohm's law of motion \eqref{Bohm} between the GRW collapses, further options besides GRWp$_5$ come to mind. Instead of \eqref{GRWp5}, we may postulate that at the time $T$ when $\psi$ collapses around $X\in\RRR^3$ with label $I$, the $I$-th particle jumps to a random position distributed with density
\be
\frac{\PPP(Q_I(T+)\in dx)}{dx} 
= \frac{\:\:\:\:\:\:\:\:\:\bigl|\psi_{T+}\bigl(Q_1(T)\ldots Q_{I-1}(T),x,Q_{I+1}(T)\ldots Q_N(T)\bigr)\bigr|^2}
{\int dx' \bigl|\psi_{T+}\bigl(Q_1(T)\ldots Q_{I-1}(T),x',Q_{I+1}(T)\ldots Q_N(T)\bigr)\bigr|^2}\,,
\ee
which is the $|\psi_{T+}|^2$ distribution conditionalized on the configuration of the other particles.

Another possibility, which we call GRWp$_6$, is that not just the $I$-th particle (the one associated with the collapse) jumps, but all particles jump.} Specifically, choose $Q(T+)$ at random with distribution $|\psi_{T+}|^2$. 

In GRWp$_6$, the $|\psi|^2$ distribution is indeed equivariant, in the sense that the configuration $Q(t)$ will always have distribution $|\psi_t|^2$. As a consequence, this theory is presumably empirically equivalent to GRWm and GRWf (in the sense that there is no experiment that could distinguish GRWp$_6$ from GRWm and GRWf). However, the particles in GRWp$_6$ do not necessarily behave in a reasonable way: For example, consider an agglomerate of $N=10^{11}$ particles and arrange a superposition of two well-separated locations, $\psi= \frac{1}{\sqrt{2}}\bigl( |\text{here}\rangle ^{\otimes N} + |\text{there}\rangle^{\otimes N}\bigr)$. Suppose the particles are all ``here'' (this happens with probability 1/2) and that the first collapse, which occurs after about 1 day, reduces $\psi$ to $|\text{there}\rangle^{\otimes N}$ (this happens independently with probability 1/2). In this case all particles jump from ``here'' to ``there.'' 

\bigskip

\textit{What this example illustrates:} Even if a theory is empirically equivalent to a reasonable theory (such as GRWm and GRWf) it need not itself be a reasonable theory.\footnote{This point is also illustrated by the following theory \cite{BellMW} that is empirically equivalent to Bohmian mechanics: Using a Schr\"odinger (i.e., non-collapsing) wave function $\psi_t$ and a particle ontology, let the configuration $Q(t)$ be random with distribution $|\psi|^2$, independently of the past.}

\subsection{MBM: Bohm-Like Trajectories From the Master Equation}
\label{sec:MBM}

The most interesting example in our list of toy theories is perhaps MBM; the abbreviation stands for ``master equation Bohmian mechanics.'' MBM resembles Bohmian mechanics in that it is deterministic and that its PO consists of particles, but at the same time it is empirically equivalent to GRWf and GRWm, as we will show in Section~\ref{sec:MBMeeGRW}.

The law of motion \eqref{Bohm} is replaced by the following equation (considered already in \cite{Bell80,dm}) using, in the role of the wave function, a density matrix $\dm$:
\begin{equation}\label{BohmDM}
    \frac{dQ_k}{dt}=v_k^{\dm}(Q_1,\ldots,Q_N)=\frac{\hbar}{m_k}
  \Im \frac{\nabla_{q_k} \langle q| \dm|q'\rangle}{\langle q|\dm |q'\rangle}
  \Big|_{q=q'=(Q_1,\ldots,Q_N)}.
\end{equation}
The density matrix $\dm$ evolves according to \eqref{M}, which we repeat here for convenience:
\begin{equation}\tag{\ref{M}}
  \frac{d \dm_t}{d t} = -\tfrac{i}{\hbar} [H,\dm_t] 
  + \lambda \sum_{k=1}^N \int d^3 x \, g_{k,x}^{1/2} \, \dm_t \, g_{k,x}^{1/2} - N\lambda \dm_t\,.
\end{equation}
We make a few comments on how these equations are to be understood. 

Eq.~\eqref{BohmDM} is the natural generalization of Bohmian mechanics to density matrices, and reduces to \eqref{Bohm} in case of a pure state $\dm = \pr{\psi}$. However, it is important to notice that the density matrix considered here is not the one that is normally regarded as the density matrix of a system, which arises by averaging $\pr{\psi}$ (in case the wave function $\psi$ is random) or by tracing out the environment of the system. The density matrix in \eqref{BohmDM}, in contrast, does not arise from averaging or partial traces but is assumed to be one of the fundamental variables of the theory. The complete description of the state is, instead of the pair $(Q,\psi)$ in Bohmian mechanics, the pair $(Q,\dm)$.

The master equation \eqref{M} is, in the GRW theories, a consequence of the GRW evolution of the wave function. This is different in MBM. In MBM there is no random wave function. In MBM, \eqref{M} holds \emph{by fiat}, not as a theorem. The defining equations of MBM---its postulates---are \eqref{M} and \eqref{BohmDM}.

\bigskip

\textit{What this example illustrates:} \x{MBM shows that the empirical content of GRWm and GRWf is compatible with a deterministic theory, and in particular does not imply wave function collapse: after all, MBM involves the master equation \eqref{M} but not literal wave function collapse as in \eqref{collapse}. If experiments will confirm the GRW deviations from quantum mechanics, then Bohmian mechanics can be modified so as to reproduce the GRW predictions. MBM also illustrates how it can make sense to speak of the density matrix of the entire universe and, more specifically, how a density matrix can be a fundamental object and part of the ontology, rather than just encoding statistical information.}

\subsubsection{Empirical Equivalence of MBM with GRWm and GRWf}
\label{sec:MBMeeGRW}

\x{We now prove the empirical equivalence. The derivation of predictions from MBM is analogous to that from Bohmian mechanics.} The analogue in MBM of the quantum equilibrium distribution $p^\psi$ described in \eqref{BMQE} is the distribution
\begin{equation}\label{MBMQE}
  p^\dm(q) = \scp{q}{\dm|q} \,.
\end{equation}
Note that $p^\dm(q) \geq 0$, and $\int_\conf p^\dm(q) \, dq = \tr \, \dm =1$. As we show in Appendix~\ref{sec:equivarianceMBM}, this distribution is equivariant in MBM as a consequence of \eqref{BohmDM} and \eqref{M}. It is therefore consistent to assume, as we do, that the configuration $Q_t$ of the world has distribution $p^{\dm_t}$ at every time $t$. And therefore, the probability at time $t$ of a certain macroscopic configuration is 
\begin{equation}
  p(S) = \int_S dq \, \langle q|\dm_t |q\rangle = \tr\bigl(\dm_t P(S)\bigr)
\end{equation}
where $S$ is the set of all microscopic configurations consistent with that macroscopic configuration, and $P(S)$ the projection operator corresponding to the set $S$, defined by 
\begin{equation}\label{Pdef}
  P(S) = \int_S dq\: \pr{q}\,.
\end{equation}
In GRW theories, $p(S)$ can be written in terms of the GRW wave function $\psi_t$ as
\begin{equation}
  \tr\bigl(\dm_t \, P(S)\bigr) = \int_\Hilbert \PPP(\psi_t \in d\phi) \, \| P(S)  \phi \|^2\,.
\end{equation}
Since $\psi_t$ is typically concentrated on a single macro-configuration, the probability distribution $\PPP(\psi_t \in d\phi)$ is typically concentrated on those $\phi$ with either $\|P(S)\phi\| \approx 0$ or $\|P(S) \phi\| \approx 1$; thus, $p(S)$ equals the probability that $\psi_t$ is (nearly) concentrated on $S$. And in this case, either the flashes of GRWf or the matter density of GRWm gives rise to the same macroscopic appearance as configurations from $S$. In other words, at any fixed time the MBM, GRWf, and GRWm worlds have the same probability distribution over the possible macro-states. 

Now empirical equivalence follows immediately: If there were an experiment which had (probably) one outcome $Z=z_1$ in MBM and another one, $Z=z_2$ in GRWm and GRWf, then at the time when the experiment is finished, the probability distribution over the macro-states would have to be different in MBM than in GRWm and GRWf, but it is not.

As a consequence of the empirical equivalence between MBM, GRWm, and GRWf, the empirical content of MBM is summarized by the GRW formalism \cite{grw3A}. (In fact, the GRW formalism was first discovered starting from MBM.) Our argument concerning empirical equivalence also exemplifies that empirical equivalence is a statement about the PO, as discussed in Section~\ref{sec:rolePO}.

\subsection{Master Equation and Matter Density}

If one can consider a version of Bohmian mechanics in which the density matrix plays exactly the role of the wave function, then why not do the same trick with the matter density and flash ontologies? 

For the matter density ontology, this would mean to postulate, in analogy to and replacing \eqref{mdef2},
\begin{equation}\label{mdefM}
  m(x,t) = \tr\bigl( \dm_t\, \widehat{M}(x) \bigr)
\end{equation}
with $\widehat{M}(x) = \sum_{i} m_i \, \delta(\widehat{Q}_i - x)$ the mass density operator, as before. Here, $\dm_t$ is taken to evolve according to the master equation \eqref{M}. We thus obtain a theory that could be called Mm, in which the fundamental objects are a density matrix (which, as in MBM, does \emph{not} represent an ensemble, or the observer's limited knowledge, but is, by postulate, a fundamental object) and the continuous matter with density $m(x,t)$. This theory, though, is very different from GRWm! It has a many-worlds character. For example, if at some initial time $\dm = \pr{\psi}$ with $|\psi\rangle = (|\text{dead}\rangle + |\text{alive}\rangle)/\sqrt{2}$ being the wave function of Schr\"odinger's cat, then after a short while the GRW function $\psi_t$ will be either $|\text{dead}\rangle$ or $|\text{alive}\rangle$, but $\dm_t$ will be $\pr{\text{dead}} + \pr{\text{alive}}$, up to a factor $\tfrac{1}{2}$. As a consequence, the $m$ field of GRWm will be either $m_\text{dead}$ or $m_\text{alive}$, but the $m$ field of Mm will be $m_\text{dead} + m_\text{alive}$, up to a factor $\tfrac{1}{2}$. Both cats are there at once, but with reduced mass (which the cats, however, do not notice). A very similar theory ``Sm'', with the unitary Schr\"odinger evolution instead of the master equation \eqref{M}, has been described in some detail in \cite{AGTZ06,mw}. It seems possible, perhaps even likely, that Sm is empirically equivalent to \x{standard quantum mechanics} \cite{mw}; if so, then for the same reasons Mm should be empirically equivalent to \x{GRWm and GRWf.}

\bigskip

\textit{What this example illustrates:} Foremost, this example illustrates the big difference, for quantum theories without observers, between a density matrix and a random wave function. Since with every probability distribution over wave functions there is associated a density matrix, and since for the purpose of computing predictions only the density matrix is relevant, it is common practice in quantum mechanics to immediately replace every probability distribution over wave functions by the density matrix. Here, however, the distinction is crucial: The deterministic $m$ function obtained from the density matrix provides a many-worlds picture of reality, whereas the random $m$ function obtained from a random wave function that is either $\dead$ or $\alive$ provides a single-world picture.

\subsection{Master Equation and Flashes}

What happens to GRWf when we replace the wave function by a density matrix? Two versions of what this might mean come to mind:

\bigskip

\noindent\textbf{MGRWf:} \x{Postulate that the universe has, at every time $t$, a density matrix $\dm_t$, whose evolution will be described below. Postulate further} that flashes occur, as in GRWf, at random times $T$ with constant rate $N\lambda$, and for a random $I\in\{1,\ldots,N\}$. In contrast to GRWf, the probability density of the flash location $X$ is given by
\be\label{MGRWfdensity}
p(x)=\frac{\PPP(X\in dx)}{dx} =  \tr\bigl( \dm_{T-}\, g_{I,x} \bigr)
\ee
instead of $\|\psi_{T-}\, g_{I,x}^{1/2}\|^2$ in GRWf. Postulate further that when a flash occurs at time $T$, the density matrix $\dm_{T-}$ changes to
\begin{equation}\label{MGRWfcollapse}
  \dm_{T+}=\frac{1}{C} g_{I,X}^{1/2}\;\dm_{T-}\; g_{I,X}^{1/2}
\end{equation}
with $C$ a normalizing constant. Between collapses, $\dm_t$ evolves unitarily 
\x{as usual. Note that if $\dm_0=|\psi_0\rangle\langle\psi_0|$ is a pure state then it will remain pure and in fact evolve according to the usual GRW dynamics. 

It follows that the joint distribution of all flashes is given by the probability measure
\begin{equation}\label{dmF}
  \PPP_{\text{MGRWf}}(F\in S)=
  \tr\bigl( \dm_0 \, \F_0(S) \bigr)
\end{equation}
with $S$ any set of flash histories and $\F_0(\cdot)$ the POVM governing the distribution of the flashes in GRWf, see \eqref{flashPOVM}. As a consequence of \eqref{dmF}, the theory MGRWf, although formulated in terms of a density matrix, is \emph{physically equivalent to GRWf!} 

Before justifying this claim, let us elucidate the notion of ``physical equivalence,'' which can be given a clear definition in the framework of PO \cite{AGTZ06}: Consider a theory $T$ (e.g., GRWf with a particular choice of $\lambda$, $\sigma$, $N$, and potential $V$) and initial data $D_0$ for $T$ (e.g., for GRWf, an initial wave function $\psi_0$); together, $T$ and $D_0$ define a probability distribution $P$ over histories of the PO after the initial time (e.g., for GRWf, a distribution over flash patterns in space-time). Now we say that the pair $(T',D'_0)$ is \emph{physically equivalent} to $(T,D_0)$ if it defines the same distribution $P$ as $(T,D_0)$. That is, two descriptions of a universe are physically equivalent if they provide the same history of the PO, or (if appropriate) the same probability distribution thereof; variables (such as $\psi$) that are not part of the PO may be different in the two descriptions.

For the purpose of comparison with MGRWf, suppose that, as the initial data $D_0$ for GRWf, we do not specify the initial wave function but instead a probability distribution $\mu$ over initial wave functions. That still defines a distribution $P$ over the PO histories: Since the joint distribution of all flashes, given $\psi_0$, is $\scp{\psi_0}{\F_0(\cdot)|\psi_0}$, the unconditional joint distribution of all flashes is
\begin{equation}
  \PPP_{\text{GRWf}}(F\in S)
  = \int\limits_{\sphere(\Hilbert)} \mu(d\psi) \, \scp{\psi_0}{\F_0(S)|\psi_0} = 
  \tr\bigl( \dm_0 \, \F_0(S) \bigr) \,,
\end{equation}
with $\dm_0$ the density matrix of the ensemble $\mu$. Since this is the same formula as \eqref{dmF}, we obtain the physical equivalence. That is, the further possibilities of initial data $\dm_0$ in MGRWf do not lead to new flash histories.

\medskip

How is MGRWf related to the master equation \eqref{M}? Since $\dm_t$ is random, it does not evolve according to \eqref{M}.} 
However, there is another density matrix, namely
\be
\tilde{\dm}_t = \EEE \dm_t\,,
\ee
where $\EEE$ means the expectation over the random flashes; $\tilde{\dm}_t$ does evolve deterministically according to the master equation \eqref{M}.

\bigskip

\textit{What this example illustrates:} First, it illustrates the concept of physical equivalence: There is no physical difference between GRWf with a random initial wave function of the universe and MGRWf, as the distribution of the PO is the same. Second, it illustrates the different roles that a density matrix can play: On the one hand, it can be part of the ontology as one of the fundamental objects, such as the random density matrix $\dm_t$. On the other hand, the other density matrix $\tilde{\dm}_t$ is a mathematical object encoding information about the probability distribution of $\dm_t$. Thus, finally, it \emph{also} illustrates that it is not necessarily appropriate to speak of ``the'' density matrix.

\bigskip

\noindent\textbf{Mf:} Another way of replacing the wave function in GRWf by a density matrix is to retain \eqref{MGRWfdensity} for the distribution of the flash location $X$ but not adopt the collapse rule \eqref{MGRWfcollapse}. Instead, let $\dm_t$ evolve deterministically according to the master equation \eqref{M}. In other words, Mf arises from MGRWf by replacing $\dm_t$ in \eqref{MGRWfdensity} by $\tilde{\dm}_t$.

This theory, too, has a many-worlds character, as the set $F$ of flashes will be the union $F=F_\mathrm{dead} \cup F_\mathrm{alive}$ of a set of flashes of a live cat and a set of flashes of a dead cat (similar to the model ``Sf'' considered in \cite{mw}). For the same reasons as for Mm, Mf is presumably empirically equivalent to GRWf and GRWm.

\bigskip

\textit{What this example illustrates:} By way of contrast with MGRWf, it illustrates the difference between the ``collapsing'' density matrix of MGRWf and the deterministic one arising from the master equation \eqref{M}. It thus illustrates why it is important that the law governing the PO be precisely formulated (since in particular, it specifies precisely which density matrix to use).

\section{Conclusions}

In this paper we have illustrated the notion of PO (primitive ontology, i.e., variables describing the distribution of matter in space and time) and its usefulness for a clean derivation of empirical predictions. We have done so by (i) describing several toy examples of theories with a PO (Section~\ref{sec:examples}), (ii) studying their predictions (Section~\ref{sec:examples}), (iii) reviewing known theories with a PO (Sections~\ref{sec:GRWmf} and \ref{sec:BM1}), and (iv) reviewing known derivations of predictions (Section~\ref{sec:rolePO}). In so doing we have illustrated the different roles that the PO, the wave function, and the density matrix have in such theories.

\appendix

\section{Proof of Equivariance in MBM}
\label{sec:equivarianceMBM}

We show that equivariance of $p^{\dm}(q) = \scp{q}{\dm|q}$ follows from \eqref{M} for the Hamiltonian \eqref{eq:H} and \eqref{BohmDM}. Before we give the formal proof, we note that the essential reason is that the non-unitary (diffusive) terms in \eqref{M} (i.e., the second and third term on the right hand side), do not contribute to the continuity equation for $p^\dm$. This can be understood by noting that a GRW collapse does not change the diagonal entries $\scp{q}{\dm_t|q}$ of the density matrix in the position representation. 

Here is the equivariance proof.
\begin{align}
  \frac{\partial}{\partial t} p^{\dm_t}(q) &=  -\tfrac{i}{\hbar} \scp{q}{[H,\dm_t]|q} 
  - N\lambda \scp{q}{\dm_t|q}  + \lambda \sum_{k=1}^N \int d^3 x \, 
  \scp{q}{g_{k,x}^{1/2} \, \dm_t \, g_{k,x}^{1/2}|q}\\
  &=  \sum_{k=1}^n \tfrac{i\hbar}{2m_k} \scp{q}{[\nabla_{q_k}^2 ,\dm_t]|q} 
  - N\lambda \scp{q}{\dm_t|q}  + \lambda \sum_{k=1}^N \int d^3 x \, 
  \scp{q}{\dm_t|q}  
  \frac{e^{-\frac{(q_k-x)^2}
  {2\sigma^2}}}{(2\pi \sigma^2)^{3/2}}\label{step2}\\
  &= \sum_{k=1}^n \tfrac{i\hbar}{2m_k} \Bigl[ (\nabla_{x_k}^2-\nabla_{y_k}^2)\scp{x}{\dm_t|y} \Bigr]_{x=y=q}
  - N\lambda \scp{q}{\dm_t|q} +N\lambda \scp{q}{\dm_t|q} \\
  &= \sum_{k=1}^n \tfrac{i\hbar}{2m_k} \Bigl[ (\nabla_{x_k}+\nabla_{y_k})\cdot(\nabla_{x_k}-\nabla_{y_k})\scp{x}{\dm_t|y} \Bigr]_{x=y=q}\\
  &= \sum_{k=1}^n \tfrac{i\hbar}{2m_k} \nabla_{q_k}\cdot\Bigl[ (\nabla_{x_k}-\nabla_{y_k})\scp{x}{\dm_t|y} \Bigr]_{x=y=q}\\
  &= -\sum_{k=1}^n \tfrac{\hbar}{m_k} \nabla_{q_k}\cdot \Im\Bigl[ \nabla_{x_k}\scp{x}{\dm_t|y} \Bigr]_{x=y=q}\\
  &=  -\sum_{k=1}^n \nabla_{q_k} \cdot (p^{\dm_t} \, v^{\dm_t}_k)\,.
\end{align}
Note that the potential $V$ in $H$ does not contribute in \eqref{step2} because
\be
\scp{q}{V\dm_t|q} =\int dq'\, \underbrace{\scp{q}{V|q'}}_{=\delta(q-q') V(q)} \scp{q'}{\dm_t|q} = V(q) \scp{q}{\dm_t|q} = \scp{q}{\dm_t V|q}\,.
\ee
Since any probability distribution $p$ on configuration space will be transported, under the flow \eqref{BohmDM}, according to the continuity equation
\begin{equation}\label{MBMconti}
  \frac{\partial p}{\partial t} = - \sum_{k=1}^N \nabla_{q_k} \cdot (p \, v^\dm_k)\,,
\end{equation}
we have that $p_t=p^{\dm_t}$ satisfies \eqref{MBMconti}, which is equivariance.

\bigskip\bigskip

\noindent\textit{Acknowledgments.}
S.~Goldstein and R.~Tumulka are supported in part by grant no.~37433 from the John Templeton Foundation. 
R.~Tumulka is supported in part by NSF Grant SES-0957568 and by the Trustees Research Fellowship Program at Rutgers, the State University of New Jersey. 
N.~Zangh\`\i\ is supported in part by INFN.

\end{document}